\documentclass[twocolumn,showpacs,preprintnumbers,amsmath,amssymb]{revtex4-1}

\usepackage{graphicx}
\usepackage{dcolumn}
\usepackage{bm}
\usepackage{textcomp}

\begin{document}

\preprint{APS/123-QED}

\title{Electron backscattered diffraction method in the analysis of deformed steel structures}

\author{Elena Pashinska, Victor Varyukhin, Anatoliy Zavdoveev, Valeriy Burkhovetskii, Valentina Glazunova}

\email{zavdoveev@fti.dn.ua}
\affiliation{Donetsk Institute of Physics and Engineering, Ukrainian
Academy of Sciences,
\\83114, R.Luxemburg str. 72, Donetsk, Ukraine
}%

\date{\today}

\begin{abstract}
The structure of low-carbon steel 
after twist extrusion is tested with 
using electron backscattered diffraction. 
It has been shown that warm twist extrusion 
results in grain refinement with conservation of 
a substantial part of high-angle boundaries, smearing 
of the texture, more homogeneous distribution of grains 
and development of the dynamic recrystallization processes.  
\end{abstract}

\maketitle

\section{Introduction}

It is well known that severe plastic deformation
 allows obtaining of materials combining high strength with 
plasticity. One of such methods is twist extrusion (TE) that is 
extruding of a prismatic billet though the die with twist channel. 
Channel geometry enables the deformed billet to preserve the size 
and the form identical to the initial state. The maximum accumulated
deformation per one pass is $ e=tg\beta$ ($\beta$ is the descent angle of the twist line) 
\cite{bvos03,bvso09}. Earlier, we analyzed the effect of TE on the 
redistribution of alloying elements in low-carbon 
construction steel and formulated general conceptions 
of morphological structure changes \cite{pdvbz11}. To obtain more 
detailed description of structure changes in low-carbon 
steels under TE, the method of automatic analysis of images 
of electron backscattering diffraction (EBSD) was used here \cite{l06}.

\section{Material and method of experiment}
As investigated material was low-carbon steel 
$20G2C$ of the following composition, mass in percent: 
$0,24 C;1,66 Mn;1,2 Si;0,14 Cr;0,24 Ni;0,01Al;0,04 S$. 
The samples were obtained by warm forging ($400$\textcelsius) and 
succeeding milling down to cross-section of $2439 mm$. 
Then the samples were annealed at $920$\textcelsius \:for $1$ hour and 
cooled in air. The extrusion was carried out with hydraulic
 press performing three passes at $P_{max} \approx 200 MPa$ and 
backpressure of $100 МPa$. Before the first pass, the 
sample was heated up to $850$ \textcelsius; before the second and 
the third pass, the heating was done up to $400$ \textcelsius; 
the temperature of the equipment was $320$ \textcelsius \: in all 
the cases. The total accumulated deformation was $e=6$.

Electron scanning microscope JSM-6490LV (JEOL, Japan) 
with special holder for bulk samples of $10$\texttimes$10$\texttimes$15$ mm in 
size was used for EBSD analysis of structure changes. The 
principle of the analysis is well known; electron beam 
scans the selected surface of the sample and Kikuchi images
 consisting of Kikuchi bands are plotted for every point. 
Every band is associated with a definite group of crystal planes. 
Using software of $HKL Chanel 5$, we established the position 
of each Kikuchi band, compared it with the theoretical data 
and calculated three-dimensional crystallographic orientation 
of the lattice. The obtained three-dimensional information was 
used for reconstruction of the microstructure.

To accumulate a representative sample, we had to analyze 
at least $1500$ grains on the specimen. Thus, the following 
procedure was applied: magnification of \texttimes$(600–700)$ was used, 
scanning area was $150$\texttimes$100 \mu$, scanning step was $500 nm$ 
(at least $10$ measuring points per a grain). The degree 
of indicating was at least $80$ percent
, the required result was not 
achieved at a lower value. 
The method allowed both 
qualitative analysis by mapping and qualitative analysis 
using statistical data. Besides, texture analysis was done. 
For more detailed structure reconstruction, the threshold of 
noise misorientation was established at $5$\textdegree. At lower threshold, 
sub-grain boundaries would be taken into account and the analysis 
of EBSD-maps would be complicated. Nevertheless, when analyzing 
distributions of grains and subgrains and drawing pole figures, 
the whole data set was accounted including misorientations below the threshold of $5$\textdegree.
\begin{figure*}
\hspace{0.06 cm}
\includegraphics [width=6 in] {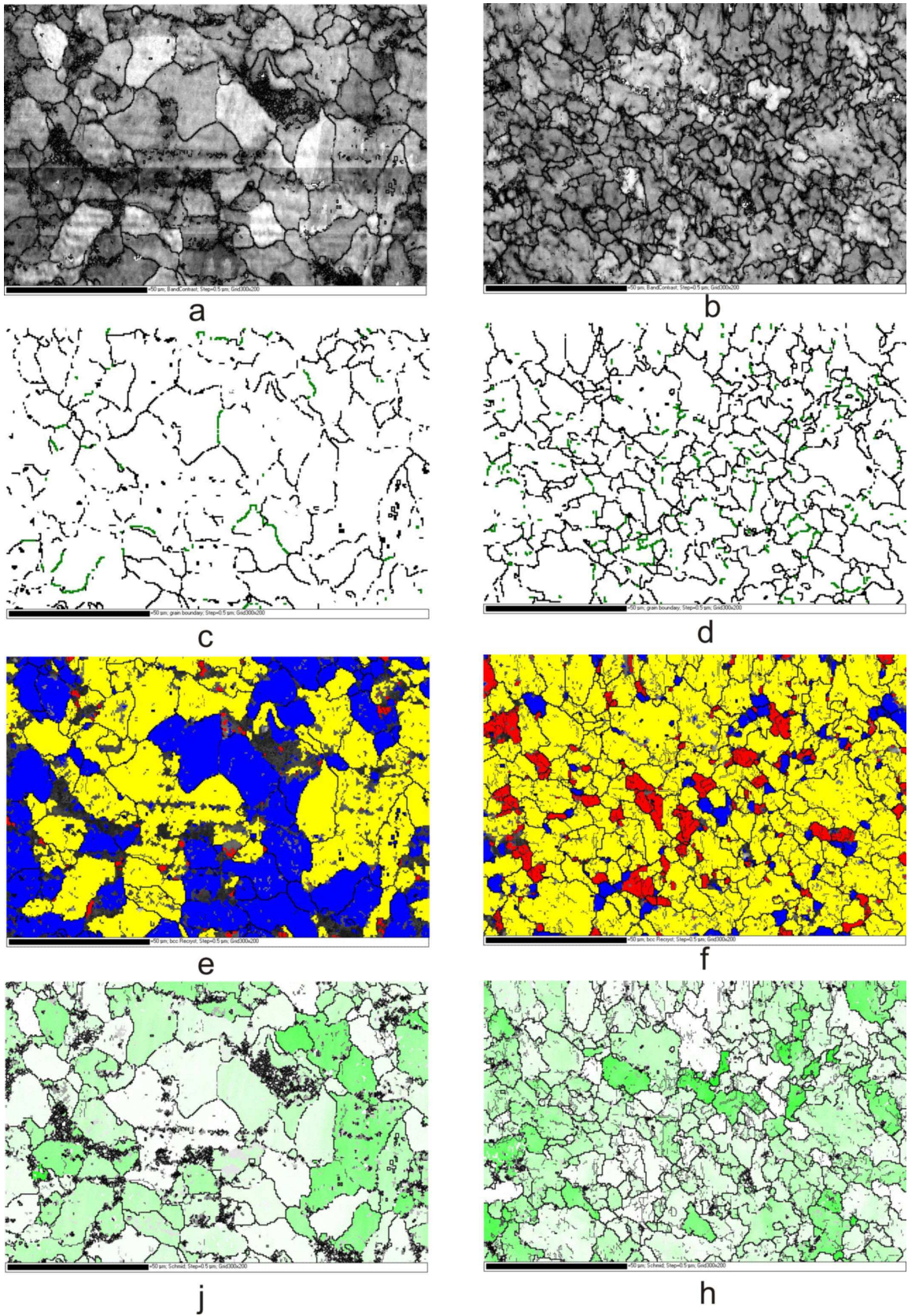}
\caption{\label{f1} Reconstruction of the microstructure of low-carbon 
steel in the initial state (a, c, e, g) and after torsional extrusion (b, d, f, h) by EBSD maps:
a,b are contrast maps; c,d are maps of grain boundary misorientation; 
e,f are maps of re-crystallization; j,h are  maps of Schmid’s factor}
\end{figure*}

\section{Results and discussion}

\subsection{Microstructure reconstruction, qualitative analysis}

For the visualization of grain morphology, contrast maps 
were created where grains were depicted in gray tones and 
grain boundaries were represented by black lines (Fig. \ref{f1}, a, b). 
Contrast maps are similar to the structure images obtained with 
the optical microscope. The maps of small-angle and high-angle 
boundaries were drawn for detailed analysis of the grain boundaries 
(Fig. \ref{f1}, c, d). Boundaries with the misorientation angle less than 
$15$\textdegree \: are colored in green and those above the given threshold are colored in black.

We have obtained also the maps characterizing the part of strained (deformed) 
and unstrained (recrystallized) grains (Fig. \ref{f1}, e, f) in steel after 
TE: unstrained recrystalized grains are marked by blue color, polygonized grains are 
marked by yellow color and deformed grains are red (These maps have been built only 
for bcc-iron. Beside bcc-iron, steel contains some fcc-iron and cementite. According 
to EBSD data, the initial composition was as follows:  
$78,5$ percent of bcc iron, $1,5$ percent of fcc iron and $20$ percent of $Fe_{3}C$;the composition in the 
deformed state was: $92$ percent of bcc iron, $2$ percent of fcc iron and $6$ percent of  $Fe_{3}C$.). 
The basic idea of these maps is that if there are no misorientations in 
the neighbor pairs of the analyzed points within a grain, the grain is 
identified as an unstrained recrystallized one. In the case when misorientations 
between two neighbor subgrains exceed $2$\textdegree, the grain is polygоnized and if they 
are above this value, the grain is a deformed one. 

Homogeneity of deformation was analyzed with the use of Schmid‘s maps. The slip 
system $<111>(110)$ was selected as the most informative one in bcc iron. 
Schmid’s factor is evaluated as \cite{l06}: $m = cos\lambda\cdotp cos\chi$, where $\lambda$ is the angle 
between the slip direction and the deformation axis, $\chi$ is the angle between the normal to the slip plane and the deformation axis. 
The maximum value of Schmid’s factor is achieved at $\chi = \lambda = 45$\textdegree. 
A macroscopic shift occurs when the shear strain in the given slip system reaches the maximum value $\tau_{0}$, that is called the critical shear strain. Shear strain $\tau$
 is related to Schmid’s factor as $\tau=\tau_{0}\cdotp m$. 
Thus, Schmid’s maps shows, what grain will be deformed earlier under uniaxial loading (these grains are colored in light, Fig. \ref{f1}, j, h). 

Correlation of ESBD-maps of different types is explicitly seen. All of them 
demonstrate grain refinement from $15 \mu$ to $5 \mu$ after TE and substantial part 
of high-angle boundaries. This fact is confirmed by data of optical and electron 
scanning microscopy. 

After warm deformation by TE, recrystallized ferrite grains are detected 
in the material (Fig. \ref{f1} e, f). They are located mostly near high-angle
 boundaries but never within large grains. Similar location of recrystallized 
volume near grain boundaries was observed in the deformed nickel alloy \cite{msp06}; 
a suggestion was made that recrystallization nuclei are formed in the most 
distorted areas of the lattice, i.e. at high-angle grain boundaries. 

Nevertheless, small recrystallized grains were found within a large polygonized 
grain. In this case, we suppose that they were formed from a few adjoining 
subgrains by defect flow toward the polygonized grain boundary inside the grain. 
The result was the reduction of defect density in the subgrain and increased misorientation 
of the boundary up to the high-angle one (Fig.\ref{f1} f). This fact means that two processes 
(dynamical polygonization and dynamical recrystallization) progress simultaneously. 
These processes can occur jointly or by turns. Thus, the analysis of EBSD maps gives 
evidences that the structure of the tested steel consists of fragmented, polygonized 
and recrystallized grains.

\subsection{Quantitative analysis}

\begin{figure}
\hspace{0.06 cm}
\includegraphics [width=3 in] {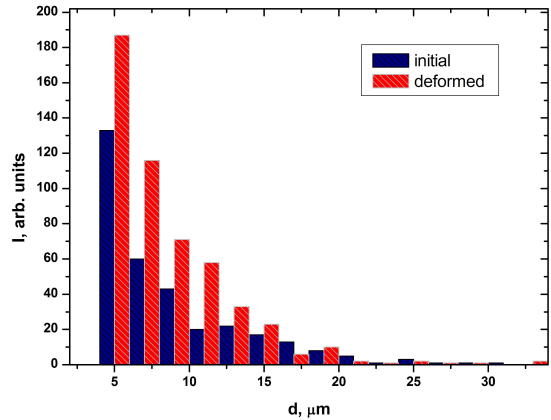}
\caption{\label{f2} Frequency distribution of grains by size}
\end{figure}

On the basis of statistical analysis of the data 
of EBSD maps, graphs of the frequency distribution 
of the grain size and misorientation angles were constructed. 
It follows from Fig. \ref{f2} that the metal structure in the initial 
state was characterized by bimodal grain distribution with the maximums 
at $13$ and $5\mu$, i.e. by explicit difference in the grain size. After TE, 
the distribution became more homogeneous with the maximum at $5\mu$ 
confirmed by EBSD-maps (see. Fig. \ref{f1}, a, b). 

\begin{figure}
\hspace{0.06 cm}
\includegraphics [width=3 in] {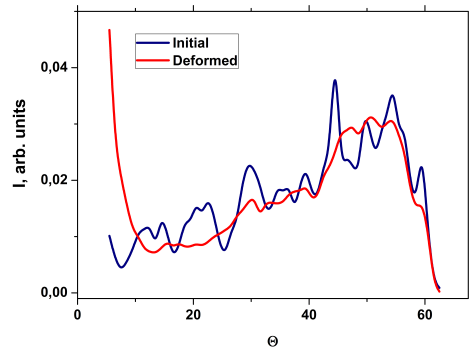}
\caption{\label{f3} Frequency distribution of misorientation angles of grain boundaries}
\end{figure}

According to the distribution of grain boundary misorientation angles 
(Fig. \ref{f3}), the specific part of high-angle boundaries in the initial 
material was about $91$ percent . This parameter was retained at high 
level ($84$ percent) in the deformed material, too. 
The formation of small-angle boundaries could be explained 
by specific features of the material. For instance, in copper 
under TE, formation of small-angle boundaries prevails at the 
initial deformation stages, but their amount is reduced at developed 
deformation \cite{vptm10}. We can suggest that high-angle grain boundaries will
 dominate in steel at accumulation of higher deformation degree. 

\begin{figure}
\hspace{0.06 cm}
\includegraphics [width=3 in] {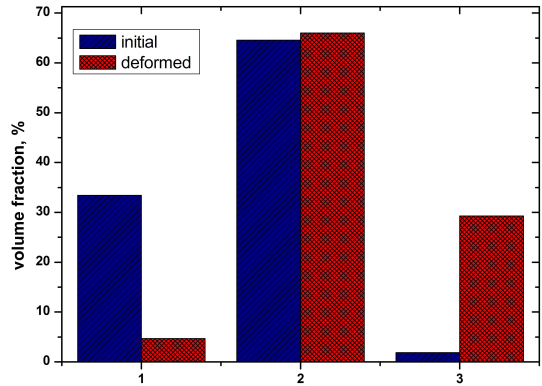}
\caption{\label{f4} Component distribution of steel microstructure:
$1$ – recrystallized grains, $2$ – polygonized grains; $3$ – deformed grains}
\end{figure}

Fig. \ref{f4} illustrates parts of different grains depending on 
the state of the material. The data were obtained by numerical 
processing of EBSD maps (Fig.\ref{f1} e,f). So, we can make a conclusion 
that the content of recrystallized grains in the initial state of 
the sample was $33,4$ percent, and this amount was reduced to $4,7$ percent
 after the deformation. As the deformation of the sample in the 
course of processing is substantial, all grains are deformed with 
succeeding passing the stages of defect density accumulation, 
fragmentation, dynamical polygonization and dynamical re-crystallization. 
Thus, occurrence of $4,7$ percent of recrystallized areas can 
be explained by active relaxation processes. The effect of 
dynamical recrystallization of steel at severe plastic deformation 
was observed in a number of experiments: for example, at ECAP \cite{dos08}, HPT\cite{gl10}. 
\begin{figure*}
\hspace{0.06 cm}
\includegraphics [width=6 in] {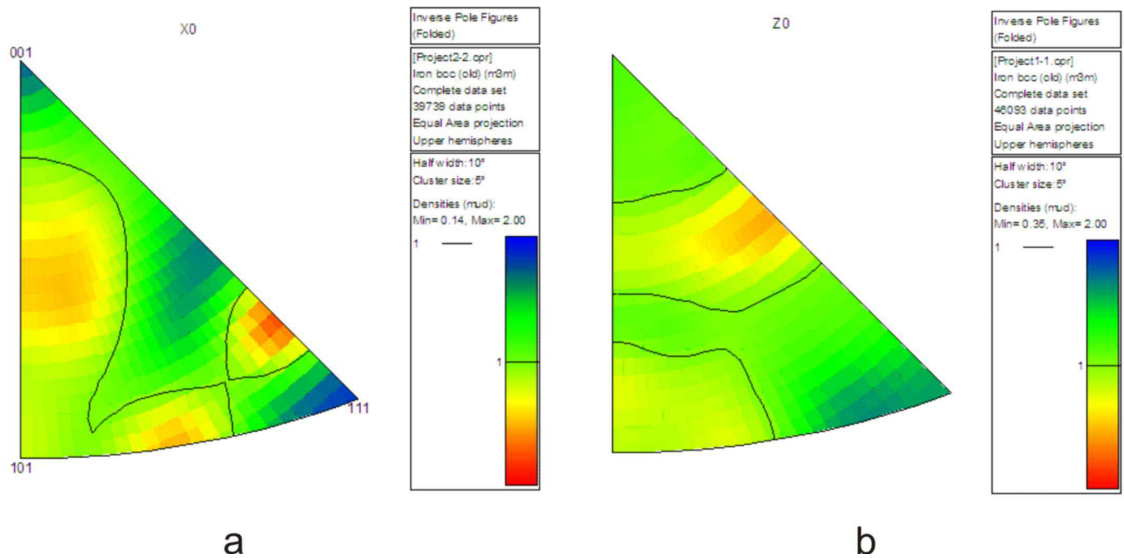}
\caption{\label{f5} Inverse pole figures of steel:  initial state (а) deformed state (b)}
\end{figure*}
\begin{figure*}
\hspace{0.06 cm}
\includegraphics [width=6 in] {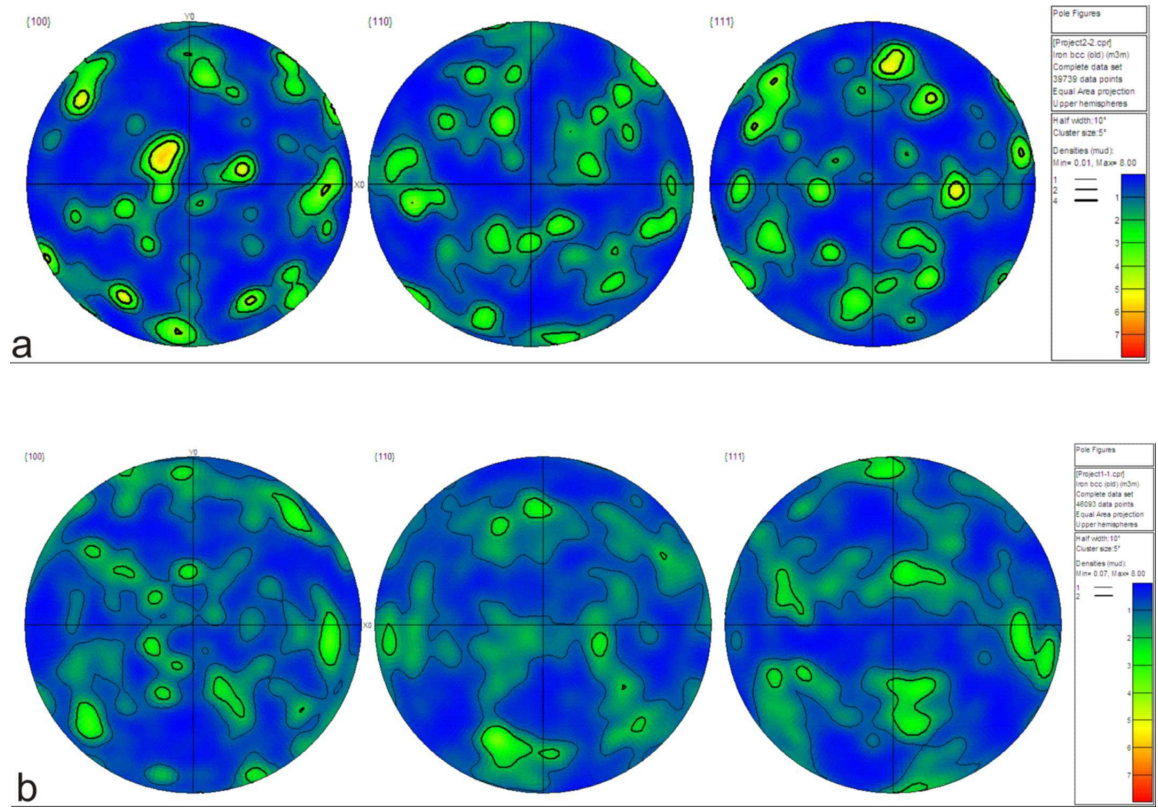}
\caption{\label{f6} Pole figures of steel:  initial state (а) deformed state (b)}
\end{figure*}

\subsection{Texture analysis}
As distinct from traditional X-ray methods used for 
the tests of large areas of a specimen, EBSD method 
allows detailed analysis of texture evolution in the 
selected zones. Figs. \ref{f5} and \ref{f6} demonstrate pole figures 
registered in the areas of about $15000 \mu^{2}$ in size. The 
analysis of inverse pole figures gives information not 
only about the presence of the texture in the sample but 
also about the mechanisms involved in the course of the 
processing. In the case of one of another prevailing pole 
density, we can make conclusions about the dominating relaxation 
mechanism (intragrain or intergrain sliding), and texture
 evolution can be related to the change of deformation mechanisms 
or development of dynamical recrystallization \cite{mmpi10}. If the maximum 
pole density is located near $<111> $and $<110>$ orientation, the 
mechanisms of twinning progressed actively within the sample \cite{ddrmj12}. 
If the intensity of the outlet of normal lines is reduced, (the texture is smearing) 
and the distribution becomes more homogeneous as in the case considered above 
(Fig. \ref{f5}), we can say about activated non-crystallographic mechanisms 
of plastic deformation.

Texture analysis cannot be accomplished without the analysis of direct 
pole figures characterizing the density of the outlet of normal 
lines of definite crystallographic planes (here these are $<111>, <110>, <100>$ planes). 
The direct pole figures confirm the fact that axial features of 
the texture are reduced in the tested steel after TE, being 
weakly expressed in general. Two factors can be associated 
with this phenomenon, these are peculiarities of the deformation 
scheme (transversal flow of metal) and relaxation processes. 
Accidentally arranged sharp texture maximums present in pole 
figures can indicate to the appearance of recrystallized grains \cite{mmpi10}. 
Because of possible ambiguity, this effect requires additional investigations.

\section{SHORT CONCLUSION}

The use of EBSD method has revealed effect 
of twist extrusion of the structure of low-carbon 
construction steel. Besides grain refinement and texture 
smearing, such treatment results in more homogeneous structure
 with a substantial part of high-angle boundaries. This 
fact is explained by peculiarities of the deformation 
scheme at twist extrusion and evolution of relaxation processes.

\begin{acknowledgments}
The authors are very grateful to Prof. 
S.V. Dobatkin for the material granted for the 
experiments, to Dr. B.M. Efros for productive discussion
 and to providing engineer B.А. Stebletsov for the aid in 
construction of the sample holder.
\end{acknowledgments}

\end{document}